\documentstyle[aps,twocolumn,prl,graphicx]{revtex}

\newcommand{\bra}[1]{\left\langle #1\right|}
\newcommand{\ket}[1]{\left| #1\right\rangle}

\newcommand{\ketbra}[2]{\left| #1\right\rangle\!\left\langle#2\right|}

\begin{document}                
\draft \wideabs{
\author{S. A. Babichev, J. Ries, A. I. Lvovsky\cite{Lvovsky}}
\address{Fachbereich Physik, Universit\"at Konstanz, D-78457 Konstanz, Germany}
\date{\today}
\title{Quantum scissors: teleportation of single-mode optical states by means of a nonlocal single photon}

\maketitle

\begin{abstract}
We employ the quantum state of a single photon entangled with the
vacuum $(\ket{1}_A\ket{0}_B-\ket{0}_A\ket{1}_B)$, generated by a
photon incident upon a symmetric beam splitter, to teleport
single-mode quantum states of light by means of the Bennett protocol.
Teleportation of coherent states results in truncation of their Fock
expansion to the first two terms. We analyze the teleported ensembles
by means of homodyne tomography and obtain fidelities of up to 99 per
cent for low source state amplitudes. This work is an experimental
realization of the quantum scissors device proposed by Pegg, Phillips
and Barnett (Phys. Rev. Lett. {\bf 81}, 1604 (1998))

\end{abstract}
 }

\paragraph{Introduction}

Quantum teleportation (QT) is the transport of an unknown quantum
state $\ket{\phi}$ over arbitrary distances by means of dual
classical and Einstein-Podolsky-Rosen (EPR) channels. To perform
teleportation, the sender, Alice, and the receiver, Bob,
prearrange the sharing of an EPR-correlated pair of particles.
Alice makes a joint measurement on her EPR particle and the source
state and sends Bob the classical result of this measurement.
Knowing this, Bob can convert the state of his EPR particle into
an exact copy of the source state. In this way neither Alice nor
Bob obtain any information about the state $\ket{\phi}$ but this
state is available at Bob's location for future use.

Although teleportation of macroscopic objects is far beyond modern
technology, quantum teleportation of microscopic states may find
its application in the observable future as a key ingredient of
quantum communication and computation. It can be used in
combination with non-deterministic computational gates to enhance
their success probability, thus making schemes involving such
gates scalable and efficient \cite{GC}. This is the role QT plays
in the recently proposed scheme for efficient quantum computation
with linear optics \cite{KLM}.

After its proposal in 1993 by Bennett {\it et al.} \cite{Bennett},
QT has been implemented experimentally on discrete- \cite{TZ,TdeM}
and continuous-variable optical states \cite{TK} as well as on
molecular spins \cite{T-NMR}. All these schemes used an EPR pair
maximally entangled in a Hilbert space which is isomorphic to the
Hilbert space of the source state. This identity allows, in
principle, exact replication of the source state by Bob.

An interesting extension of the Bennett protocol arises, however, if
the source state lives in the Hilbert space of {\it higher} dimension
than the EPR pair. In this case all terms of the source state
associated with the dimensions beyond that of the EPR pair will be
``cut off" from the teleported state. This is known as the {\it
quantum scissors} (QS) effect first described theoretically in 1998
by Pegg, Phillips, and Barnett \cite{QS} and implemented
experimentally in the present work.

In the heart of our teleportation experiment there is an EPR pair
implemented by a nonlocal single photon state
$\ket{\Psi^-}=\frac{1}{\sqrt{2}}(\ket{1}\ket{0}-\ket{0}\ket{1})$
which is generated when a single photon $\ket{1}$, incident upon a
symmetric beam splitter, entangles itself with the vacuum $\ket{0}$.
It is remarkable that our EPR ensemble is formed by just one
particle; yet this is a maximally entangled state in the
two-dimensional Hilbert space defined by basis vectors $\ket{0}$ and
$\ket{1}$.

We apply the state $\ket{\Psi^-}$ to teleport arbitrary
single-mode quantum states of the electromagnetic field that
belong to the Hilbert space of infinite dimension. If a random
quantum state is given by $\ket{\phi}=\sum a_n\ket{n}$ in the
number (Fock) state basis, the scissors effect will truncate the
above series, leading to the output of a form $\ket{\phi_{\rm
out}}=a_0\ket{0}+a_1\ket{1}$. In simpler words, the higher number
terms cannot reach Bob because there is never more than one photon
in the original EPR state $\ket{\Psi^-}$.

In our actual experiment, the role of the source ensemble was played
by a coherent state $\ket{\alpha}$. Since this state has an infinite
number of terms in its Fock expansion, it is well suitable for
demonstrating quantum scissors; on the other hand it is readily
available from the source laser. Alice's Bell-state measurement is
performed by overlapping the source state and her share of the EPR
state on a beam splitter and measuring the number of photons in each
output (Fig.~1). Events in which the detector $D_1$ registers exactly
one photon while $D_2$ registers zero photons correspond to the
two-mode state $\ket{\Psi^-}_{12}$ entering Alice's apparatus in
modes 1 and 2. If this is the case, Bob's share of the EPR state
(mode 3) is in the state $\ket{\phi_{\rm out}}$ so no additional
manipulations are required from Bob to complete the QT protocol.
Restricting to these events, we perform a homodyne measurement on the
teleported ensembles in order to characterize them and determine the
teleportation fidelity.

Full implementation of the scissors protocol requires, in particular,
highly efficient single-photon detectors capable of determining the
number of incident photons. Although such detectors are currently
being developed \cite{Yamamoto}, they are not widely available.
Fortunately, the protocol exhibits surprisingly good fidelity even
with regular, non-discriminating single-photon detectors as long as
the amplitude of the source state is sufficiently small
\cite{BarnettPRA,ImotoPRA}.

\begin{figure}[tbp]
\begin{center}
\includegraphics[width=0.4\textwidth]{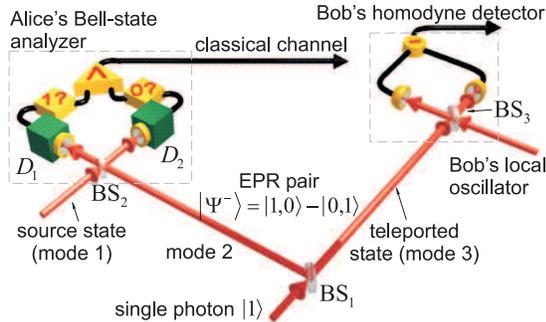}
\caption{\label{Scheme}Conceptual scheme of the experiment.
BS$_i$, beam splitters; $D_i$, single photon detectors.}
\end{center}
\end{figure}

The QS protocol finds its direct application as an integral part
of the single-rail version of the linear optical quantum computer
recently proposed by Lund and Ralph \cite{LundRalph}. Apart from
the quantum gate efficiency enhancement discussed earlier, this
technique is useful for preparing quantum bits of arbitrary value.
Although conditional preparation of superpositions
$a_0\ket{0}+a_1\ket{1}$ has been reported earlier
\cite{Catalysis,Steinberg01}, the present scheme is highly
efficient, employs only passive optics and can be extended to
multi-photon ensembles \cite{GenQS}.


In a related experiment, Lombardi {\it et al.} recently used the
nonlocal photon for partial teleportation of a single-photon qubit
encoded in a dual optical channel. One part of the entangled
source state was teleported to Bob by means of a Bell measurement
while the other was transported to Bob directly \cite{DeM}. The
main difference between Ref. \cite{DeM} and this work is that the
former required a direct quantum channel between Alice and Bob to
complete the transfer of a quantum state, whereas the latter is a
straightforward implementation of the Bennett protocol
\cite{Bennett} in the form of quantum scissors \cite{QS}.

We note that our teleportation scheme is of {\it a priori} nature in
the context of the discussion \cite{T9}, i.e. teleportation events
need not be postselected according to the result of Bob's
measurements. Whenever Alice obtains a positive result of the Bell
measurement, it is known that the teleportation has been successful
and the teleported state is available for future use.

\paragraph{Theory}
The practical implementation of the QS protocol requires a source
of single photons in a pure optical mode matching that of the
source ensemble. We solve this task by means of conditional
measurement on a down-converted pair: the single photon state
$\ket{1}$ is prepared in the signal channel of the down-converter
when its counterpart is detected in the idler \cite{Fock,MMpaper}.
Theoretical treatment of the scissors protocol in this
configuration has been elaborated by \"Ozdemir {\it et al.}
\cite{ImotoPRA,ImotoXXX}, so here we only present a brief
overview.

Preparation of the single photon state is imperfect: dark counts
of our trigger detector ($D_T$) and optical losses result in a
statistical mixture of one photon and no photon instead of a pure
$\ket{1}$ state. The ensemble entering the first beam splitter is
therefore
\begin{eqnarray}\hat{\rho}_{\ket{1}}=\eta_{\ket{1}}\ketbra{1}{1}+(1-\eta_{\ket{1}})\ketbra{0}{0},\end{eqnarray}
$\eta_{\ket{1}}$ being the preparation efficiency.

The density matrix of the EPR pair used for teleportation
$\hat{\rho}^{\rm epr}$ can be found by applying the beam splitter
transformation operator
\begin{eqnarray}
\nonumber\hat{B}\ket{m,n}&=&\sum_{j,k=0}^{m,n}
\sqrt{\frac{(j+k)!(m+n-j-k)!}{m!n!}}
\left(\begin{array}{c}m\\j\end{array}\right)
\left(\begin{array}{c}n\\k\end{array}\right)\\ \lefteqn{\times
(-1)^k 2^{-(n+m)/2} \ket{j+k,\,m+n-j-k}}.
\end{eqnarray}
 to the incident combination of $\hat{\rho}_{\ket{1}}$ and the vacuum:
\begin{eqnarray}
\hat{\rho}^{\rm epr}=\hat{B}\left(
\ket{0}\bra{0}\otimes\hat{\rho}_{\ket{1}}\right)\hat{B}^\dag.
\end{eqnarray}

The source coherent state and one of the EPR ``particles" enter
Alice's apparatus where they undergo further transformation via
another beamsplitter. After this transformation, the density
matrix of the 3-mode ensemble can be written as
\begin{eqnarray}
\hat\rho_{123}=\hat{B}_{12}\left(\ket{\alpha}_1\bra{\alpha}_1\otimes\hat{\rho}_{23}^{\rm
epr}\right)\hat{B}^\dag_{12},
\end{eqnarray}
where the subscripts refer to the optical modes according to Fig.~1.

The first two modes of $\hat\rho_{123}$ are subjected to
measurements via single-photon detectors. A non-discriminating
detector of quantum efficiency $\eta_{\rm SPD}$ is described by
the following positive operator-valued measure (POVM):
\begin{eqnarray}
\hat{\Pi}^{\rm no-click}&=&\sum_{n=0}^\infty (1-\eta_{\rm SPD})^n
\ketbra{n}{n}\nonumber\\ \hat{\Pi}^{\rm
click}&=&\hat{1}-\hat{\Pi}^{\rm no-click}.\label{POVM}
\end{eqnarray}
This measurement leads to a collapse of $\hat{\rho}_{123}$ projecting
it in the event of a ``click" in detector $D_1$ and ``no click" in
detector $D_2$ upon the following non-normalized ensemble in Bob's
channel:
\begin{eqnarray}\label{rho_out}
\hat{\rho}_{\rm out}={\rm Tr}_{12}\
(\hat{\rho}_{123}\,\hat\Pi^{\rm click}_1\,\hat\Pi^{\rm
no-click}_2).
\end{eqnarray}
The probability of a teleportation event is given by $p_{\rm
tel}={\rm Tr}(\hat{\rho}_{\rm out})$.

Imperfect spatial, spectral or temporal mode matching between
Alice's share of the nonlocal single photon and the source state
$\ket{\alpha}$ leads to partial distinguishability and more
classical-like behavior, reducing the teleportation fidelity. In
case of a complete mode mismatch, the behavior of the system is
fully described by a semiclassical model in which photons act like
particles with no wave properties. Each beam splitter distributes
the incident photons randomly into the output channels. The
correlated photon number distribution in the three modes can be
calculated according to the laws of classical statistics. From
this distribution we infer the probability $p_{\rm tel}^{\rm sc}$
of the positive Bell measurement outcome as well as the
conditional probability $p_{\rm out}^{\rm sc}$ that Bob's mode
contains a photon. The ensemble received by Bob can then be
expressed as a density matrix
\begin{equation}
\hat{\rho}_{\rm out}^{\rm sc}=\left( {\matrix{{1-p_{\rm out}^{\rm
sc}}&0\cr 0&p_{\rm out}^{\rm sc}\cr }} \right).
\end{equation}


In the actual case of partial mode matching, the output ensemble
is a mixture of those calculated via classical and semiclassical
models
\begin{equation}\label{rho_bob}
\hat{\rho}^{\rm Bob} = M\,p_{\rm out}\,N[\hat{\rho}_{\rm
out}]+(1-M)p_{\rm out}^{\rm sc}\,N[\hat{\rho}_{\rm out}^{\rm sc}],
\end{equation}
where $M$ is the mode matching factor {\cite{MMpaper} and
$N[\hat\rho]$ denotes normalization.

Imperfections in the homodyne detection of the teleported state
such as poor mode matching between the local oscillator and the
signal, linear losses, inefficient photodiodes or imperfect
balance can all be modeled by a single beam splitter with one
empty input in the signal beam with a reflectivity $\eta_{\rm HD}$
(generalized Bernoulli transformation) \cite{MMpaper,leon}.


\paragraph{Experimental apparatus}
 The setup for preparing the single-photon Fock state was
the same as in our previous experiments \cite{Catalysis,Fock}. A
82-MHz repetition rate train of 1.6-ps pulses generated by a
Spectra-Physics Ti:Sapphire laser at 790 nm was frequency doubled
and directed into a beta-barium borate crystal for
down-conversion. The latter occurred in a type-one
frequency-degenerate, but spatially non-degenerate configuration.
The single-photon detector $D_T$, placed into the idler channel of
the down-converter, detected photon-pair creation events and
triggered all further measurements.

Pulses containing conditionally-prepared photons entered the optical
arrangement shown in Fig.~\ref{Scheme}, which had to be maintained
interferometrically stable throughout the experimental run. The
coherent source state $|\alpha\rangle$ and the local oscillator for
homodyne detection were provided by the master Ti:Sapphire laser.
These two modes had to be matched, spatially and temporally, to the
respective modes of the EPR pair. To this end, we modeled the single
photon by a classical wave as described in \cite{Fock,MMpaper}. The
mode matching was then optimized by maximizing the visibility of the
interference fringes observed in the beam splitter outputs. The
visibility value provided a basis for a ballpark estimation of the
mode matching factor $M$.

Further knowledge of the experimental parameters was gained
through an auxiliary tomography measurement in which the ensemble
arriving to Bob was characterized without conditioning on Alice's
results. This ensemble is a statistical mixture of states
$\ket{0}$ and $\ket{1}$ with the single-photon fraction equal to
$\eta_{\ket{1}}\eta_{\rm HD}/2$. We found $\eta_{\ket{1}}\eta_{\rm
HD}\approx 0.49$

All single-photon detectors used were from Perkin-Elmer, SPCM-AQ
series, with quantum efficiencies (including the filtering optics) of
about $\eta_{\rm SPD}\approx 0.5$. The homodyne measurement of the
teleported state was conditioned upon detectors $D_1$ and $D_T$
firing and the detector $D_2$ not firing. The digital logic employed
featured rigorous synchronization control of the photon count events
with respect to each other and to the master laser pulses. This
helped us reduce the dark count contribution to a negligible level.


The time-domain homodyne detector used for characterizing the
teleported state was described in \cite{FockHD}. For each value of
$\alpha$ approximately 20000 events were collected. The phase of the
local oscillator was varied with a piezoelectric transducer. The
acquired data was used to calculate the density matrix
$\hat{\rho}_{\rm out}^{\rm exp}$ of the teleported ensemble by means
of the quantum state sampling method \cite{leon}. The teleportation
fidelity was then evaluated as

\begin{equation}\label{F}
F=\bra{\alpha}\hat{\rho}_{\rm out}^{\rm exp}\ket{\alpha}.
\end{equation}

\paragraph{Results and discussion}
For conceptual verification of the teleportation protocol we
performed a measurement run in which we varied the phase of the
source state instead of the local oscillator. From the classical
point of view, this action should not affect the optical field
observed by Bob and therefore its quadrature statistics should
remain constant. Yet we observed the optical phase of the
teleported ensemble vary in accordance with that of the source
(Fig.~\ref{Results}). This result is readily explained by quantum
mechanics: by changing the source state phase Alice changes the
conditions of the measurement performed on one of the members of
the EPR pair. This has a nonlocal effect on the other member which
is observed by Bob in his homodyne measurement.

\begin{figure}[tbp]
\begin{center}
\includegraphics[width=0.38\textwidth]{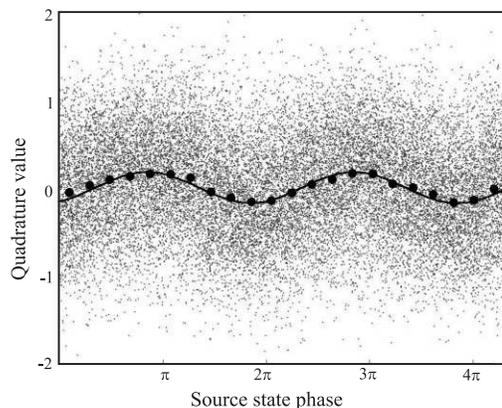}
\caption{\label{Results}Quadrature noise distribution of the
teleported state measured with a balanced homodyne detector as a
function of the source state phase. }
\end{center}
\end{figure}

Fig.~\ref{Fidelity} shows the teleportation fidelity determined
experimentally along with the theoretical fit calculated according
to Eqs.~(\ref{rho_out})--(\ref{F}). There were three fitting
parameters: quantum efficiencies $\eta_{\ket{1}}$ and $\eta_{\rm
HD}$ of the single photon preparation and the homodyne detection,
respectively, and the mode matching factor $M$. By fitting these
parameters with fixed $\eta_{\rm HD}\cdot\eta_{\ket{1}}$ we found
$\eta_{\rm HD}= 0.54 $, $\eta_{\ket{1}}=0.9$, $M= 0.56$. Note that
the value of $\eta_{\rm HD}$ includes not only the homodyne
detector efficiency {\it per se}, but also the mode matching of
Bob's ensemble with the local oscillator.

Along with the data pertinent to the actual experiment,
Fig.~\ref{Fidelity} also shows the behavior of the fidelity factor
for the idealized quantum-mechanical model with number
discriminating detectors and the semiclassical particle model
discussed above. All three models exhibit similar qualitative
behavior. If the source state is vacuum, a photon detected by
Alice must originate from the EPR pair, so Bob receives no
photons. The ensemble arriving at Bob's station is in the vacuum
state, and the teleportation fidelity is perfect. For high
$\alpha$, the input state has almost vanishing vacuum and
single-photon terms, the only components of the truncated
teleported ensemble. The teleported ensemble is then practically
orthogonal to the source state, and the fidelity is low.

Our experimentally measured fidelity is always higher than that
predicted semiclassically, showing the importance of quantum
nonlocal effects. A remarkable feature is that for low values of
$\alpha$ the value of $F$ is very high, up to 99 per cent. To our
knowledge, this is the highest fidelity ever achieved in
experimental QT.

\begin{figure}[tbp]
\begin{center}
\includegraphics[width=0.4\textwidth]{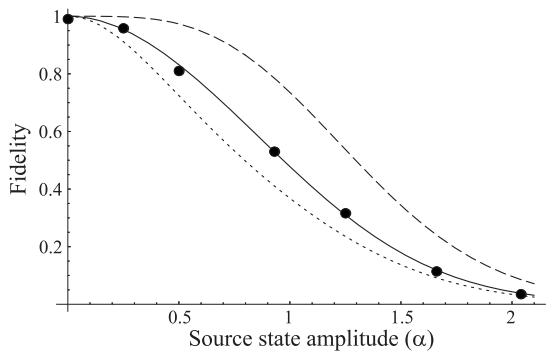}
\caption{\label{Fidelity}Teleportation fidelity as a function of
the amplitude $\alpha$ of the coherent source state. Shown are the
experimental data and the theoretical fit (solid line), the
idealized quantum mechanical model with number discriminating
detectors (dashed line) and the semiclassical particle model
(dotted line). }
\end{center}
\end{figure}

\paragraph{Conclusion}
We reported an experimental realization of quantum scissors, i.e.
teleportation of single-mode optical ensembles using the nonlocal
single photon state as the EPR pair. The teleported state was
examined by homodyne measurement and the fidelity was found to be
well above the classical limit. Since we did not postselect the
teleportation events according to Bob's results, this experiment
is of {\it a priori} nature. To our knowledge, this is the first
QT experiment in which the Bell measurement was done in a
discrete, and the characterization of the teleported state in a
continuous basis.

In perspective we plan to improve our teleportation fidelity by
using number discriminating photon detectors \cite{Yamamoto}.
Another possibility would be to extend the QS protocol to
synthesize arbitrary truncated superpositions of Fock states
$a_0\ket{0}+...+a_n\ket{n}$ \cite{GenQS,WelschPRA}. The nonlocal
single photon $|0,1\rangle+|1,0\rangle$ is worth further
investigation from the point of view of quantum nonlocality
\cite{nonlocality}.

This work was sponsored by the Deutsche Forschungsgemeinschaft and
the Optik-Zentrum Konstanz.


\end{document}